\documentclass[prl,reprint,superscriptaddress,,nofootinbib]{revtex4-2}
\pdfoutput=1
\usepackage{amsfonts, amssymb, amsmath}
\usepackage{mathrsfs}
\usepackage[utf8]{inputenc}
\usepackage[russian,english]{babel}
\usepackage{float}

\usepackage{enumitem}
\setlist{nosep, leftmargin=*} 

\usepackage{color}
\definecolor{dark-gray}{gray}{0.20}
\definecolor{gray}{gray}{0.30}
\definecolor{light-gray}{gray}{0.80}
\definecolor{dark-red}{rgb}{0.7,0,0}
\definecolor{dark-green}{rgb}{0.1,0.4,0}
\definecolor{dark-blue}{rgb}{0.3,0.3,0.7}
\definecolor{light-blue}{rgb}{0.8,0.8,1}
\definecolor{blue}{rgb}{0,0,1}
\definecolor{red}{rgb}{1,0,0}
\definecolor{green}{rgb}{0,1,0}

\usepackage{hyperref}
\hypersetup{
	colorlinks=true,
	linkcolor=dark-blue,
	citecolor=dark-red,
	urlcolor=dark-blue,
	linktoc=page
}

\def\cI{{\cal I}}

\def\cM{{\cal M}}
\def\cN{{\cal N}}

\def\cC{{\cal C}}

\def\cZ{{\cal Z}}

\def\i{{\rm i}}

\newcommand{\be}{\begin{equation}}
\newcommand{\ee}{\end{equation}}
\newcommand{\ba}{\begin{aligned}}
\newcommand{\ea}{\end{aligned}}
\newcommand{\bea}{\begin{eqnarray}}
\newcommand{\eea}{\end{eqnarray}}

\newcommand{\mathd}{\mathrm{d}}
\newcommand{\mathe}{\mathrm{e}}
\newcommand{\mathi}{\mathrm{i}}

\newcommand{\e}{\epsilon}
\newcommand{\lam}{\lambda}
\newcommand{\BV}{\mathbb{V}}

\newcommand{\BC}{\mathbb{C}}

\newcommand{\WPL}{\mathbb{WP}^1_{a,b}}

\begin{document}

\title{Indices of M5 and M2 branes at finite $N$ \\from equivariant volumes, and a new duality}

\author{Kiril Hristov}

\affiliation{Faculty of Physics, Sofia University, J. Bourchier Blvd. 5, 1164 Sofia, Bulgaria}
\affiliation{INRNE, Bulgarian Academy of Sciences, Tsarigradsko Chaussee 72, 1784 Sofia, Bulgaria}
\affiliation{School of Mathematics and Physics, University of Queensland, St Lucia, Brisbane, Queensland 4072, Australia}

\begin{abstract}
We study supersymmetric indices of the 6d $(2,0)$ theory of $N$ M5‑branes on toric Sasaki–Einstein five‑manifolds. Embedding the background into a local toric Calabi‑Yau four‑fold and equivariantly integrating the anomaly polynomial yields a finite‑$N$ Cardy‑limit formula in terms of equivariant characteristic classes. Separately, using equivariant constant maps in topological string theory and higher‑derivative supergravity, we derive a finite‑$N$ proposal for the superconformal, twisted, and spindle indices of $N$ M2‑branes probing arbitrary toric Calabi‑Yau four‑folds. The M2‑brane partition functions depend on the same combination of equivariant classes as the M5 result. Motivated by this match, we generalize the M2/M5 duality recently discussed in \cite{Chen:2026fpe} to an infinite class of M2‑brane theories by exchanging the worldvolume and transverse geometries of the two brane systems.
\end{abstract}
\date{\today}
\maketitle


\section{Introduction}
\label{sec:intro}
Branes are central objects in string theory and holography, and dualities between distinct brane systems often uncover unexpected relations among quantum field theories in different dimensions. A recent proposal relating the superconformal indices of M2- and M5-brane theories \cite{Chen:2026fpe} (inspired from studies of giant graviton expansions~\cite{Gaiotto:2021xce,Imamura:2021ytr,Hayashi:2024aaf})—which can be viewed as a novel extension of standard holographic correspondences—calls for a clearer conceptual and geometric understanding. In this work, we provide an extension of this duality in the perturbative regime of the partition functions (at finite $N$), and clarify its geometric origin using equivariant methods.~\footnote{See also \cite{Batrachenko:2002pu} for a related proposal.}

On the M5-brane side, we focus on Sasaki-Einstein (SE$_5$) indices. We build on the well-established relation between anomaly polynomials and Cardy limits of even-dimensional superconformal field theories (SCFTs) on compact backgrounds: supersymmetric partition functions are strongly constrained by anomalies, and equivariant integration localizes their evaluation to fixed-point data (see, e.g., \cite{Kim:2012ava,Assel:2015nca,Bobev:2015kza,Brunner:2016nyk,Nahmgoong:2019hko,Hosseini:2020vgl,Ohmori:2021dzb,Hosseini:2021fge,Martelli:2023oqk,Cassani:2024tvk}). In this work, together with a forthcoming paper \cite{Hristov:2026}, we extend this framework beyond compact spaces to non-compact toric geometries. Utilizing the formalism in \cite{Martelli:2023oqk,Colombo:2023fhu,Cassia:2025aus}, we treat local toric Calabi–Yau (CY) manifolds as natural $(d+2)$-dimensional extension spaces whose codimension-two boundary defines the physical $d$-dimensional background. This perspective is natural from the viewpoint of anomalies, since the anomaly polynomial is a $(d+2)$-form, and it provides a unified geometric origin for the SE$_5$ squashing parameters.

On the M2-brane side, we exploit recent progress in equivariant topological strings, where constant-map contributions capture partition functions at finite $N$ \cite{Cassia:2025aus,Cassia:2025jkr}. In the perturbative regime, there is strong evidence that the (squashed) $S^3$ partition function is entirely determined by equivariant characteristic classes of the geometry (see, e.g., \cite{Kapustin:2009kz,Drukker:2010nc,Hama:2011ea,Marino:2011eh,Hatsuda:2013oxa,Nosaka:2015iiw,Hatsuda:2016uqa,Chester:2021gdw,Geukens:2024zmt,Kubo:2024qhq,Bobev:2025ltz,Hristov:2026zjh}). Combining these results with constraints from higher-derivative supergravity \cite{Bobev:2021oku,Hristov:2021qsw,Hristov:2022plc,Hristov:2024cgj}, we extend this structure to twisted and superconformal indices (on $S^1 \times S^2$) of M2-brane theories, as well as to twisted and anti-twisted spindle indices (on $S^1 \times \WPL$).

Taken together, the anomaly-based equivariant integration on the M5 side and the constant-map sector of equivariant topological strings on the M2 side provide a unified geometric framework that explains, from first principles, the structure underlying the proposed M2/M5 duality, and furnish non-trivial perturbative evidence for it at finite $N$.

Concretely, from an M‑theory perspective, M5‑branes split 11d spacetime into six worldvolume directions $\cM_6$ and five transverse flat directions $\cN_5=\mathbb{R}^5$. The M5 anomaly polynomial is naturally defined on an eight‑dimensional extension $X_8$ with $\partial X_8=\cM_6$, which we take to be a toric CY four-fold (to be derived). As shown in \cite{Martelli:2023oqk}, the normal bundle can be described equivariantly on $Z_4=\mathbb{C}^2$. Thus an M5 partition function is specified by the choice of spaces
\be
\text{M5:}\qquad X_8(\parallel)\times Z_4(\perp)\ .
\ee
M2‑branes instead occupy three worldvolume directions $\cM_3$ with eight transverse directions $\cN_8$. For backgrounds preserving at least eight supercharges we take the cone $\cN_8=C(SE_7)$; denote a resolution of this cone by $X_8$, with first Chern class $c_1(X_8)=0$ (again CY). In the near‑horizon limit the parallel directions are asymptotically Euclidean AdS$_4$, so topologically the worldvolume can be modelled by $Z_4=\mathbb{C}^2$. Hence for M2‑branes
\be
\text{M2:}\qquad Z_4(\parallel)\times X_8(\perp)\ .
\ee
The condition for both brane systems to preserve supersymmetry is that the total equivariant first Chern class vanishes,~\footnote{Here and in the Appendix, we denote the equivariant upgrade of a given quantity with a superscript $\mathbb{T}$.}
\be
\label{eq:8}
	c_1^\mathbb{T} (X_8) = c_1^\mathbb{T} (Z_4)\ .
\ee
Both M2 and M5 systems admit canonical (fixed $N$) and grand-canonical (conjugate $\mu$) ensembles, see \cite{Gautason:2025plx}. Their relation is
\vspace{-2mm}
\be
\label{eq:ensembles}
\cZ(\mu) = \sum_{N} \cZ(N)\, e^{\mu N}\ , \quad \cZ (N) = \frac1{2 \pi i} \int {\rm d} \mu\, \cZ(\mu)\, \mathe^{-\mu N}\ .
\ee
Using this, we present evidence for a duality between M5 partition functions in the canonical ensemble and M2 partition functions in the grand‑canonical ensemble (or vice‑versa):~\footnote{We drop dimension indices: in what follows $X$ is a complex toric four‑fold (8 real dimensions), $Y$ a toric three‑fold (6 real dimensions), $Z$ a toric two‑fold (4 real dimensions), and $L$ an arbitrary toric Sasakian space (5 real dimensions).}
\be
\label{eq:newduality}
\cZ_{\rm M5}\big(N_{\rm M5}; X(\parallel)\times Z(\perp)\big) \leftrightarrow \cZ_{\rm M2}\big(\mu_{\rm M2}; Z(\parallel)\times X(\perp)\big)\ ,
\ee
with the explicit map between $N_{\rm M5}$ and $\mu_{\rm M2}$ discussed later. The crucial novel ingredient – beyond the exchange of ensembles already present in the original proposal of \cite{Chen:2026fpe}  – is the interchange of parallel and transverse spaces. Further details (e.g., “thermal” equivariant parameters fixed to a constant) are given below; we present concrete evidence for the case:
\be
\label{eq:spaceX}
X=\mathbb{C}\times Y\ , \qquad Z=\mathbb{C}^2\ ,
\ee
where $Y$ is an arbitrary local toric Calabi–Yau threefold.~\footnote{Note that our heuristic identification of $Z$ is not relied upon in explicit calculations, but we keep it to illustrate the general idea.} This yields an infinite family of examples, generalising the special case $Y=\mathbb{C}^3$ discussed in \cite{Chen:2026fpe}. We emphasise that our results rely solely on classical equivariant geometry, capturing only perturbative finite $N$ corrections on both sides. Proving the correspondence at the exact quantum level remains open. We now detail our M5 and M2 brane calculations and assumptions; the complete duality statement appears at the end.

\section{M5-branes: anomaly polynomial and SE$_5$ indices}
\label{sec:se-index}

We first focus on the index of the 6d $(2,0)$-theory of $N$ coincident M5-branes on a (squashed) five‑dimensional SE manifold $L$, with background $\cM_6 = S^1 \times L$ (see \cite{Qiu:2013pta,Qiu:2013aga,Schmude:2014lfa,Qiu:2014oqa,Alday:2015lta}). The Cardy limit of the SE$_5$ index is,
\begin{align}
\label{eq:1}
	\cI^{L}_{\rm M5} (N; \omega, \Delta) := -2 \pi \i \int_{\BC \times Y} {\rm P}^\mathbb{T}_8\ ,
\end{align}
where the equivariant upgrade of the $8$-form anomaly ${\rm P}_8$ is integrated on $X = \BC \times Y$, a smooth resolution of $C(S^1) \times C(L)$. Since $\cM_6$ is a product of Sasakian manifolds, $X$ automatically satisfies the Calabi-Yau condition $c_1(X_8)=0$, and $\partial X = \cM_6$ by construction. Here $\omega_{\hat \imath}$ are equivariant parameters on $Y$ (squashing parameters on $L$, \cite{Martelli:2023oqk,Cassia:2025aus}), and $\Delta_{1,2}$ are associated with the Cartan of $SO(5)$.\\

{\bf Equivariant integration}\\
The anomaly of the 6d $(2,0)$ theory, associated to a simply-laced Lie algebra $G$, is the eight-form \cite{Witten:1996hc,Freed:1998tg,Harvey:1998bx,Intriligator:2000eq,Yi:2001bz,Ohmori:2014kda}
\be
\label{eq:7}
\begin{split}
	{\rm P}_8 &= \frac{h_G d_G+r_G}{24}\, p_2 (Z) \\
+& \frac{r_G}{48}\, \left( \frac14\, (p_1  (Z) - p_1 (X) )^2 - p_2 (Z) - p_2 (X) \right)\ ,
\end{split}
\ee
where $h_G, d_G, r_G$ are the dual Coxeter number, dimension and rank of the group $G$, respectively.
Please consult the Appendix for details on the characteristic classes appearing above, and their equivariant upgrades. As stated in the introduction, we will take the normal bundle to be topologically $\BC^2$, following \cite{Martelli:2023oqk}. In this case $G=SU(N)$, so $(h_G d_G + r_G) = N^3-1$ and $r_G = N-1$.

\begin{table}[H]
\centering
\setlength{\tabcolsep}{6pt}
\begin{tabular}{ c || c | c | } \hline
\multicolumn{1}{|c||}{\textbf{CY manifold}} & ${\rm dim}_\BC$ & \text{$\e$-parameters}  \\ \hline \hline
\multicolumn{1}{|c||}{$X (\parallel)$}  & $4$ & $\e_i$ \\ \hline
\multicolumn{1}{|c||}{$Y \subset X$} & $3$ & $\omega_{\hat \imath}$ \\ \hline
\multicolumn{1}{|c||}{$Z (\perp) = \BC^2$} & $2$ & $\Delta_{1,2}$ \\ \hline
\end{tabular}
\caption{Spaces relevant for the M5-brane description}
\label{tab:linear}
\end{table}

Before specializing to the concrete space $X$ in \eqref{eq:spaceX}, we consider it to be an arbitrary toric CY four-fold, with $\e_i$ its equivariant parameters. Motivated by all known examples, though lacking a complete derivation (see e.g.\ \cite{Cassani:2024tvk}[Section 2] for a discussion), we assume that supersymmetry preservation corresponds to \eqref{eq:8}. After a short computation (cf.\ \eqref{eq:a10}--\eqref{eq:a11}), this  implies
\be
\label{eq:mainconstr}
	 k^X_1 (\e) = \sum_i \e_i = k_1^Z (\Delta) = \Delta_1 + \Delta_2\ ,
\ee
with the shorthand functions $k_p$ (proportional to equivariant Chern numbers) defined in \eqref{eq:defk}.
Utilizing this constraint, we can present a rather compact expression for the equivariant integral of the anomaly polynomial $\rm P_8$ on the toric manifold $X$:
\begin{widetext}
\be
\label{eq:generalintegral}
\begin{split}
	 \int_X  {\rm P}_8^\mathbb{T} &= \frac{(N^3-1)}{24}\, (\Delta_1 \Delta_2)^2\, C_X (\e) +  \frac{(N-1)}{24}\, \Big( k^X_1 (\e) k^X_3 (\e) - (\Delta_1 \Delta_2) k^X_2(\e)- k^X_4 (\e) \Big)\, C_X (\e) \\
	&= \frac{C_X}{24}\, \Big( (N^3-1)\, (k_2^Z)^2 + (N-1)\, \left( k_1^X k_3^X-k_2^Z k_2^X - k_4^X \right) \Big)\ ,
\end{split}
\ee
\end{widetext}
where the separate contributions appearing in \eqref{eq:7} have been evaluated in \eqref{eq:shorthandc2andc3} and \eqref{eq:a13}, and $C_X$ denotes the equivariant volume at vanishing K\"ahler parameters. The expressions above are subject to the constraint \eqref{eq:mainconstr}, which allows equivalent re‑expressions. All quantities are uniquely determined by the generating function of equivariant intersection numbers, $\BV(\lam,\e)$, defined in the Appendix and computed explicitly in many examples in \cite{Martelli:2023oqk,Cassia:2025aus} and references therein. This calculation is the main technical result of the paper; we now turn to its physical interpretation.\\

{\bf Sasaki-Einstein indices}\\
We specialize the auxiliary space  to the direct product space $X= \BC \times Y$, with $Y$ a smooth resolution of the cone $C(L)$ over a five-dimensional Sasakian manifold $L$, \eqref{eq:spaceX}. There is substantial evidence in the literature, see \cite{Nahmgoong:2019hko,Ohmori:2021dzb,Cassani:2024tvk} and the relation with direct computations in \cite{Honda:2019cio,ArabiArdehali:2019tdm,Kim:2019yrz,GonzalezLezcano:2020yeb,Goldstein:2020yvj,Amariti:2021ubd,Cassani:2021fyv,ArabiArdehali:2021nsx}, that the equivariant integral computed on $X$ reproduces, as a function of $N$, the Cardy limit of the corresponding partition function.  For the present purposes, in line with the standard normalization in e.g.\ \cite{Chen:2026fpe}, we assert that~\footnote{By Cardy limit we mean the ``high temperature'' limit used in \cite{Kantor:2019lfo,Amariti:2021ubd,Cassani:2021fyv,ArabiArdehali:2021nsx}, sometimes referred to as the Cardy limit on the second sheet, as opposed to the one on the first sheet considered in e.g.\ \cite{Chang:2019uag}. The present version of the Cardy limit may also be related to the Casimir energy, see \cite{Ohmori:2021dzb}.}
\be
	\lim_\text{Cardy}\Big[ - \log \cZ_{M5} (S^1 \times L)  \Big]= \cI_{M5}^L\ ,
\ee
with the latter quantity defined in \eqref{eq:1}. 

In order to explicitly evaluate $\cI_{M5}^L$ from the general formula \eqref{eq:generalintegral}, we further assign $\e_0$ as the unique equivariant parameter of $\BC$ and use $\omega_{\hat \imath}$ for $Y$ (with conjugate parameters $\gamma$), $\e_i = \{\e_0, \omega_{\hat \imath}\}$. It can be shown that, \cite{Cassia:2025aus}
\be
\label{eq:12}
	\BV_{\BC \times Y} = \frac{\mathe^{\lambda^0 \e_0}}{\e_0}\, \BV_Y (\gamma, \omega)\ ,
\ee
which allows further simplification of many of the above formulas. The special equivariant parameter $\e_0$ relates to the size of the ``thermal'' circle (i.e.\ the Euclideanized and compactified time direction) of the real space, on which the theory lives. The equivariant parameter is therefore fixed to take a constant value, which conventionally we choose as~\footnote{Here we follow the convention of \cite{Cassani:2024tvk}, which also fixes the overall prefactor in \eqref{eq:1}.}
\be
	\e_0 = -2 \pi \i\ ,
\ee
which in turn implies the following supersymmetry constraint stemming from \eqref{eq:mainconstr}:
\be
\label{eq:14}
k_1^{\BC \times Y} = \e_0 + k_1^Y (\omega) = -2 \pi \i + \sum_{\hat \imath} \omega_{\hat \imath} = \Delta_1 + \Delta_2\ . 
\ee
We further observe the following identities that follow identically from \eqref{eq:12}:~\footnote{For related proposals that replace Pontryagin classes by an additional ``thermal'' term, see \cite{Nahmgoong:2019hko,Ohmori:2021dzb,Chen:2026fpe}; our derivation shows these replacements arise effectively from the inclusion of $\e_0$.}
\be
\label{eq:simplificationsforCxY}
\begin{split}
	C_{\BC \times Y} &= \frac{C_Y (\omega)}{\e_0}\ , \quad k_2^{\BC \times Y} = \e_0\, k_1^Y (\omega) + k_2^Y (\omega)\ , \\
	k_3^{\BC \times Y}&= \e_0\, k_2^Y (\omega) + k_3^Y (\omega)\ , \quad k_4^{\BC \times Y} = \e_0\, k_3^Y (\omega)\ .
\end{split}
\ee 

We then find the following result for the Cardy limit of the Sasakian index:
\begin{widetext}
\be
\label{eq:CardySasaki}
	\cI^{S^1 \times L}_{M5}  (N; \omega, \Delta) = \frac{(N^3-1)}{24}\, (\Delta_1 \Delta_2)^2\, C_Y 
+ \frac{(N-1)}{24}\, \left( k^Y_1 k^Y_3 - 2 \pi \i k^Y_1 (k^Y_2 - \Delta_1 \Delta_2) - k^Y_2 (4 \pi^2 + \Delta_1 \Delta_2)  \right)\, C_Y\ ,
\ee
\end{widetext}
where, for brevity, the explicit dependence of $k^Y_p$ and $C_Y$ on $\omega$ is suppressed.\\

{\bf Examples.}\\
The most studied example is $L = S^5$, see \cite{Cvetic:2005zi,Chow:2007ts,Hosseini:2018dob,Bobev:2023bxl,Bobev:2025xan} for the dual black holes in AdS$_7$, which we can generalize to $L = S^5 / \mathbb{Z}_q$. In this case $Y = \BC/\mathbb{Z}_q$, or more precisely a resolution of it. We then find three equivariant parameters, $\omega_{1,2,3}$, and
\be
\begin{split}
	C_{\BC/\mathbb{Z}_q} & = (q\, \prod_{{\hat \imath}=1}^3 \omega_{\hat \imath})^{-1}\ , \\
k_2^{\BC/\mathbb{Z}_q} = & \sum_{{\hat \imath}<{\hat \jmath}} \omega_{\hat \imath} \omega_{\hat \jmath}\ ,\quad  k_3^{\BC/\mathbb{Z}_q} = q^2\, \prod_{\hat \imath} \omega_{\hat \imath}\ ,
\end{split}
\ee
see \cite{Cassia:2025jkr}[Section 4.2], noting that the Euler number is $\chi (\BC/\mathbb{Z}_q) = q$, the total number of fixed points. It is straightforward to check that in the case $q=1$, the expression \eqref{eq:CardySasaki} reproduces precisely the answer in \cite{Chen:2026fpe}.

Another typical example is the resolved conifold, $Y = \cC$, i.e.\ the resolution of the cone over $T^{1,1}$. The results are again readily available, with total of four (redundant) equivariant parameters $\omega_{1,2,3,4}$:
\be
\begin{split}
	C_\cC &= \frac{\sum_{{\hat \imath}=1}^4 \omega_{\hat \imath}}{(\omega_1+\omega_3) (\omega_2+\omega_3) (\omega_1+\omega_4) (\omega_2+\omega_4)}\ , \\
 k_2^\cC &= \omega_1 \omega_2 + \omega_3 \omega_4+ \sum_{{\hat \imath}<{\hat \jmath}} \omega_{\hat \imath} \omega_{\hat \jmath}\ ,\quad  k_3^{\cC} = 2\, (C_\cC)^{-1}\ ,
\end{split}
\ee
such that the Euler number is $\chi (\cC) = 2$. See \cite{Hristov:2026uxs} for the $Y^{p,q}$ and $L^{p,q,r}$, where the explicit expressions are slightly longer.

\section{M2-branes: $S^3$ partition function and spindle indices}
\label{sec:spindle}
We now turn to the $N$ coincident M2-brane system, where the transverse space $X$ is an arbitrary local toric CY four‑fold, preserving eight supercharges asymptotically. Flat space $X = \BC^4$ gives the ABJM theory \cite{Aharony:2008ug}; more general $X$ lead to flavoured versions of 3d SYM or ABJM theories and their quiver generalisations \cite{Benini:2009qs}. Unlike the M5 case, we now adopt a bulk perspective and aim to reproduce the finite‑$N$ partition functions of these quiver theories using only the topological data of $X$. This is achieved via the equivariant volume (see Appendix) together with equivariant topological string constant map terms \cite{Cassia:2025aus,Martelli:2023oqk,Colombo:2023fhu}. Details of the calculation appear in \cite{Cassia:2025jkr}, while a condensed review is given in \cite{Hristov:2026zjh}.
\begin{table}[H]
\centering
\setlength{\tabcolsep}{6pt}
\begin{tabular}{ c || c | c | } \hline
\multicolumn{1}{|c||}{\textbf{CY manifold}} & ${\rm dim}_\BC$ & \text{$\e$-parameters}  \\ \hline \hline
\multicolumn{1}{|c||}{$Z (\parallel) = \BC^2\ , \BC \times \WPL$} & $2$ & $\nu_0, \nu$ \\ \hline
\multicolumn{1}{|c||}{$X (\perp)$}  & $4$ & $\tilde \e_i$ \\ \hline
\end{tabular}
\caption{Spaces relevant for the M2-brane description}
\label{tab:linear}
\vspace{-1mm}
\end{table}

{\bf $S^3$ partition function}\\
The main result of \cite{Cassia:2025jkr} for the three-sphere partition function with a squashing parameter $\nu$  is given by,~\footnote{For a more immediate comparison with the M5 results, we use rescaled parameters with respect to \cite{Cassia:2025jkr}: $\tilde \e_\text{here} =\i \pi\, (1+b_\text{there}^2)\, \e_\text{there}, \nu_\text{here} = 2 \pi \i\, b^2_\text{there}$. Note also that we switch off baryonic symmetries from the outset, see \cite{Hosseini:2025mgf}.}
\be
\label{eq:Airys3}
	\cZ_{M2} (N; S^3_\nu) \simeq \text{Ai}  \Big( \frak{C}^{-1/3} (\tilde \e, \nu) (N - \frak{B} (\tilde \e, \nu)) \Big)\ , 
\ee
with
\begin{align}
\begin{split}
& \frak{C}(\tilde \e, \nu) = -\frac{\nu^2}{2}\,  C_X (\tilde \e)\ , \\ 
\frak{B}(\tilde \e, \nu) &= \frac{ C_X (\tilde \e)}{24}\, \Big( k_4 (\tilde \e) + 2 \pi \i \nu k_2 (\tilde \e) - (2 \pi \i + \nu)^2\, \frac{k_3 (\tilde \e)}{ k_1(\tilde \e)} \Big)\ ,
\end{split}
\end{align}
where the equality in \eqref{eq:Airys3} holds up to constant prefactors (in $N$) and non-perturbative corrections. We have labeled by $\tilde \e_i$ the equivariant parameters of the transverse space $X$ in order not to confuse them with their analogs entering the M5-brane description.  In this case we find the following supersymmetric constraint: 
\be
\label{eq:21}
		k^X_1 (\tilde \e) = \sum_i \tilde \e_i = 2 \pi \i + \nu =: \nu_0 + \nu = k^Z_1\ ,
\ee
which again formally coincides with \eqref{eq:8}. We also defined the constant parameter $\nu_0$, which we interpret as the "thermal" parameter of the M2-brane, part of the internal (or worldsheet) $Z$ manifold. The Airy function of first kind, ${\rm Ai}$, appearing above, has the following integral representation and asymptotic expansion,
\be
\label{eq:Airyexp}
	{\rm Ai} (z) = \frac1{2 \pi \i} \int_C {\rm d}\tilde \mu\, \mathe^{\frac{\tilde \mu^3}{3} -z\, \tilde \mu}  \sim \frac{\mathe^{-\frac23 z^{3/2}}}{2 \sqrt{\pi} z^{1/4}}\ ,
\ee
where we only presented the first term in the asymptotic expansion, which will be useful later, and the contour is defined by $C: e^{-\i \pi/3}\, \infty \longrightarrow e^{\i \pi/3}\, \infty$. In the grand-canonical ensemble, \eqref{eq:ensembles}, we then find using \eqref{eq:21}
\begin{widetext}
\be
\label{eq:muensembleS3}
\begin{split}
	-\log \cZ_{M2} (\mu; S^3_\nu) &\simeq - \frac{\mu^3}{24 \pi^2}\,(2 \pi \i \nu)^2\, C_X (\tilde \e) + \frac{\mu}{24}\, \Big( k^X_1 (\tilde \e) k^X_3 (\tilde \e) - (2 \pi \i \nu) k^X_2 (\tilde \e) - k^X_4 (\tilde \e) \Big)\, C_X (\tilde \e) \\
	& = \frac{C_X}{24}\, \Big( -\frac{\mu^3}{\pi^2}\, (k_2^Z)^2 + \mu\, \left( k_1^X k_3^X-k_2^Z k_2^X - k_4^X \right) \Big)\ .
\end{split}
\ee
\end{widetext}
up to constant and non-perturbative corrections in $\mu$. Strikingly, the above formula and \eqref{eq:generalintegral} involve identical characteristic class combinations, despite arising from entirely unrelated calculations. Beyond strongly suggesting the announced M2/M5 duality, this comparison hints at a relation between topological string theory and anomaly polynomials, raising the question of whether other fundamental string objects are similarly related.\\

{\bf Effective 4d supergravity}\\
To further explore the finite‑$N$ result, we make one additional assumption. The near‑horizon M2‑brane geometry is AdS$_4 \times SE_7$; reduction on the seven‑manifold yields an effective 4d supergravity. Switching on all equivariant parameters requires a truncation that includes all isometries of the internal manifold. Although every such SE$_7$ space admits a consistent truncation \cite{Gauntlett:2007ma} (see also \cite{Cassani:2012pj}), a general ansatz retaining all relevant Kaluza-Klein (KK) modes is still missing. Nonetheless, using the explicit supergravity dual of the squashed three‑sphere \cite{Hristov:2021qsw,Hristov:2022plc}, we can fully constrain this putative effective supergravity and predict all higher‑derivative (HD) corrections from the finite‑$N$ expression in \eqref{eq:Airys3}.

As discussed in detail in \cite{Hristov:2021qsw}, the Lagrangian of gauged HD 4d $\mathcal{N}=2$ supergravity with $n_V$ physical abelian vector multiplets is determined by the prepotential
\be
\label{eq:prepot}
	F(X^I; A_\mathbb{W},  A_\mathbb{T}) = \sum_{m, n = 0}^\infty \, F^{(m,n)} (X^I)\, (A_\mathbb{W})^m\, (A_\mathbb{T})^n\ ,  
\ee
which depends on the off‑shell vector multiplet complex scalars $X^I$ ($I = 0, \dots, n_V$) and the composite scalars $A_\mathbb{W}, A_\mathbb{T}$ that generate the higher‑derivative Weyl‑squared \cite{Bergshoeff:1980is} and T‑log \cite{Butter:2013lta} invariants, respectively. Supersymmetry imposes that each holomorphic $F^{(m,n)}(X^I)$ is homogeneous of degree $2(1-m-n)$.

Given a theory defined by the above prepotential, the HD sugra action for the squashed sphere boundary is given by, \cite{Hristov:2021qsw,Hristov:2022plc}
\be 
	I_{S^3_\nu} (\varphi^I, \nu) = \frac{2 \pi}{\nu}\, F \left(\varphi^I; (2\pi\i-\nu)^2, (2\pi\i+\nu)^2 \right)\ ,
\ee
under the constraint $ \sum_i \varphi^i = 2\pi \i + \nu$. Comparing directly with \eqref{eq:Airys3} and assuming supergravity only reproduces the (negative) exponent in the Airy expansion \eqref{eq:Airyexp}, we predict that the effective higher‑derivative Lagrangian from compactification on a SE$_7$ space (whose resolved cone is the local toric CY manifold $X$) is,~\footnote{Mapping to supergravity requires identifying equivariant parameters with supergravity scalars; different conventions also shift index numbering and positioning.}
\be
\hspace{-2mm}
\label{eq:central}
	F= \frac{\sqrt{2}\, \i}{3\pi\, \sqrt{C_X (X^I)}}\,  \left( N_\chi - k_{\mathbb{W}} (X^I) A_{\mathbb{W}} -  k_{\mathbb{T}} (X^I) A_{\mathbb{T}} \right)^{3/2}\ ,
\ee
with the functions $k_{\mathbb{W}, \mathbb{T}}$ given by
\be
\hspace{-3mm}
	 k_{\mathbb{W}} := -\frac1{96}\, k^X_2\, C_X \ , \quad k_{\mathbb{T}} := \frac{1}{96} \left( k^X_2 -4\, \frac{k^X_3}{k_1^X}\right) C_X\ ,
\ee
both homogeneous of degree $-2$ as required by supersymmetry. We have introduced the shifted rank
\be
	N_\chi := N - \frac{\chi (X)}{24}\ ,
\ee
still assumed large in the supergravity approximation, such that a Taylor expansion of \eqref{eq:central} fits in the form \eqref{eq:prepot}. Supergravity imposes stricter constraints on the prepotential than the form of \eqref{eq:Airys3} alone—in particular, we cannot freely reexpress terms using \eqref{eq:21} without breaking the correct homogeneity pattern of \eqref{eq:prepot}. Hence the existence of an acceptable prepotential \eqref{eq:central} already provides a nontrivial check that our consistent truncation assumption is meaningful.\\

{\bf Spindle and sphere indices}\\
The effective supergravity lets us apply supergravity localization \cite{BenettiGenolini:2023kxp,BenettiGenolini:2023ndb,Hosseini:2019iad}. In particular, the HD gluing result of \cite{Hristov:2021qsw,Hristov:2024cgj} yields general predictions for topologically twisted and superconformal indices on the sphere and on spindles ($\WPL$) \cite{Benini:2015noa,Benini:2015eyy,Kim:2009wb,Imamura:2011su,Kapustin:2011jm,Inglese:2023wky,Colombo:2024mts}. The resulting gluing rule for black spindles gives the supergravity action as
\be
\hspace{-1mm}
\begin{split}
	I_{S^1\times \WPL} &= \frac{2 \pi}{\nu} \Big[ F \left(\varphi^{I,+}; (2\pi\i-\nu)^2, (2\pi\i+\nu)^2 \right) \\
	-& \sigma\, F \left(\varphi^{I,-}; (2\pi\i+\sigma\, \nu)^2, (2\pi\i-\sigma\, \nu)^2 \right) \Big]\ ,
\end{split}
\ee
with co-prime integers $a, b$ corresponding to conical deficit angles, \cite{Ferrero:2020laf,Ferrero:2020twa}, and
\be
	\varphi^{I, \pm} := \varphi^I \mp \frac{n_i}{2 a b}\ , \qquad \sigma:= \frac{b}{|b|}\ ,
\ee
where we take $a>0$; $\sigma = 1$ corresponds to topological twist, $\sigma = -1$ to anti‑twist, see \cite{Ferrero:2021etw}. The chemical potentials $\varphi^I$ couple to electric charges $q_i$, and $\nu$ to angular momentum $J$. Magnetic charges and chemical potentials satisfy the supersymmetric conditions~\footnote{In this case, deriving the supersymmetry conditions directly from \eqref{eq:8} is less straightforward, as the twist (or anti-twist) conditions on the spindle mix the $R$-symmetry bundle with the tangent bundle; see \cite[Section 4.2]{Martelli:2023oqk} for a careful explanation.}
\be
\label{eq:31}
	\sum_i n_i = a+b\ , \qquad \sum_i \varphi^i = 2 \pi \i + \frac{a-b}{2 a b}\, \nu\ .
\ee

Translating this prediction into geometric variables and repackaging into Airy functions yields
\begin{widetext}
\be
\begin{aligned}
	\text{twist}: \quad \cZ_{M2}^{\sigma = 1}   \simeq & {\rm Ai} \Big[ \left(\frak{C} (\tilde \e^+, \nu) \right)^{-1/3}\, (N - \frak{B} (\tilde \e^+, \nu)) \Big] \times {\rm Bi} \Big[ \left( \frak{C} (\tilde \e^-, \nu) \right)^{-1/3}\, (N - \frak{B} (\tilde \e^-, - \nu)) \Big]\ , \\
	\text{anti-twist}: \quad \cZ_{M2}^{\sigma=-1} &  \simeq {\rm Ai} \Big[ \left(\frak{C} (\tilde \e^+, \nu) \right)^{-1/3}\, (N - \frak{B} (\tilde \e^+, \nu)) \Big] \times {\rm Ai} \Big[ \left( \frak{C} (\tilde \e^-, \nu) \right)^{-1/3}\, (N - \frak{B} (\tilde \e^-, \nu)) \Big]\ ,
\end{aligned}
\ee
\end{widetext}
where ${\rm Bi}$ is the Airy function of the second kind, and with $\frak{C}, \frak{B}$ as in \eqref{eq:Airys3}, as well as
\be
	\tilde \e_i^{\pm} := \tilde \e_i \mp \frac{n_i}{2 a b}\, \nu\ , \qquad  \sum_i \tilde \e_i = 2 \pi \i + \frac{a-b}{2 a b}\, \nu\ ,
\ee
extending the prediction of \cite{Cassia:2025jkr}.

We can recover the two sphere indices in the limit $a = |b| = 1$: for $\sigma = +1$ we recover the (refined) topologically twisted index (TTI), while for $\sigma = -1$ we find the (generalized) superconformal index (SCI). The TTI admits an unrefined limit of exactly vanishing equivariant (or refinement) parameter $\nu$:
\be
	\cZ_{M2}^\text{TTI} = \cZ_{M2}^{\sigma = 1} (a = b = 1, \nu = 0)\ .
\ee
The SCI instead admits a limit where all magnetic charges are vanishing, $n_i = 0$, allowed by \eqref{eq:31} since $a+b =0$. One then simply finds the SCI to be the square of the three-sphere partition function (at the level of precision we are working with)
\bea
\hspace{-3mm}
\begin{split}
	\cI_{M2}^{SCI} (N; \tilde \e, \nu) &:=	-\log Z^{\sigma = -1}_{M2} (a=b=1, n_i = 0) \\
 &= - 2\,  \log \cZ_{M2} (N; S^3_\nu)\ ,
\end{split}
\eea
under the constraint $\sum_i \tilde \e_i = 2 \pi \i  + \nu$.

\section{A new duality}
Although we already noted the similarity between \eqref{eq:generalintegral} and \eqref{eq:muensembleS3}, we can now make the relation outlined in \eqref{eq:newduality} more explicit. As suggested there, we take the same manifold on both sides, here $X = \mathbb{C} \times Y$. Using the simplifications in \eqref{eq:simplificationsforCxY}, the grand-canonical ensemble of the SCI for M2 branes on $\mathbb{C} \times Y$ becomes
\vspace{-1mm}
\bea
\hspace{-5mm}
\begin{split}
	& \cI^\text{SCI}_{M2} (\mu; \tilde \e_0, \omega, \nu) \simeq   \frac{C_Y }{24\, \tilde \e_0}\, \Big[ 8\, \mu^3\, \nu^2 \\
+2\, \mu & \left(  ( (\tilde \e_0)^2 - 2\pi \i \nu + \tilde \e_0\, k_1) k_2 + (k_3 - 2 \pi \i \nu \tilde \e_0) k_1 \right) \Big]\ ,
\end{split}
\eea
where we dropped the tilde on $\omega$ but kept $\tilde \e_0$ as the constraints \eqref{eq:14} and \eqref{eq:21} are not equivalent. Up to constant terms in $N$ (outside the precision of the calculation), the relation between partition functions is
\be
\hspace{-2mm}
	\cI^\text{SCI}_{M2} (\mu = -\frac{N \Delta_1}{2}; \tilde \e_0 = -\Delta_1,  \omega, \nu = \Delta_2) = \cI^{S^1 \times L}_{M5} (N; \omega, \Delta)\ ,
\ee
where the identification can be permuted between $\Delta_1$ and $\Delta_2$ by symmetry. This agrees with the suggestion of \cite{Chen:2026fpe} (up to a factor of $2$) and now holds for an arbitrary toric three‑fold $Y$, realizing explicitly the proposed \eqref{eq:newduality}. The general form of this relation is reminiscent of the AGT-duality, \cite{Alday:2009aq}, given one starts from a 12-dimensional viewpoint, see \cite{Hristov:2026uxs}, relating to the anomaly inflow idea, \cite{Witten:1996hc,Freed:1998tg,Harvey:1998bx}. Note that this analysis cannot distinguish on the left hand side between the superconformal index and twice the $S^3$ partition function; finer checks are needed to see if the match persists at the full quantum level. Finally, the proposed duality naturally gets extended via the equivariant CY$_4$/CY$_3$ correspondence recently discussed in \cite{Hristov:2026zjh}, which relates the $S^3$ partition function (and by the extension of this work, also the SCI) of M2‑brane theories on more general toric CY$_4$ spaces to those on $\mathbb{C} \times$CY$_3$.

\section*{Acknowledgements}
I would like to thank Canberk Sanli and all authors of \cite{Chen:2026fpe} for discussions and for motivating the present work. I am also very grateful to Luca Cassia and Ali Mert Yetkin for collaborations on related topics. I am supported in part by the Bulgarian NSF grant KP-06-N88/1.

\bibliographystyle{apsrev4-2}
\bibliography{refs.bib}

\clearpage

\appendix{}

\onecolumngrid

\setcounter{section}{1}
 \setcounter{equation}{0}
 
\begin{center}\textbf{APPENDIX}\end{center}

\section{\large 
\textbf{
A.  Equivariant volume and characteristic classes}}
\label{sec:equi}

A complex toric manifold $X$ can be realized as the K\"ahler quotient $X = \BC^n // U(1)^r$, where the complex dimension is $d := n-r$. A diagonal $U(1)^n$ acts on $\BC^n$, and the embedding of the quotient $U(1)^r \subset U(1)^n$ is specified by the integer charge matrix $Q^a_i$ (GLSM charges), with $i = 1, \dots, n$ and $a = 1, \dots, r$. Working equivariantly with respect to the full $U(1)^n$, we upgrade the symplectic form to an equivariant form $\omega \rightarrow \omega^\mathbb{T}$ satisfying $({\rm d} + \e_i \iota_{u^i})\, \omega^\mathbb{T} = 0$, where $u^i$ generate the $U(1)^n$ action. We further parametrize the Kähler form by redundant parameters $\omega = \omega_\lam$, with formal variables $\lam^i$ that overparametrise the physical K\"ahler moduli $t^a$ via $t^a = Q^a_i \lam^i$, with summation over repeated indices.

Following \cite{Martelli:2023oqk} and \cite{Cassia:2025aus}, we define the generating function of equivariant intersection numbers,
\be
\label{eq:evol-JK}
\BV_X(\lam,\e) := \int_X \mathe^{\omega_\lambda^\mathbb{T}} = \oint_\text{JK} \prod_{a=1}^r\frac{\mathd\phi_a}{2\pi\mathi} \frac{\mathe^{x_i\lam^i}}{\prod_{i=1}^n x_i}\ , \qquad  x_i = \e_i+\phi_aQ^a_i \in H^\ast_{U(1)^n}(X)\ ,
\ee
where in the last step above we used the relation between equivariant integration, fixed-point localization and Jeffrey-Kirwan (JK) residue respresentation for the toric manifolds,
\be
	\int_X \leftrightarrow \sum_\text{fixed pts} \leftrightarrow \text{JK-residues}\ .
\ee
Two practical approaches exist for computing $\BV_X$ above: fixed point localization (emphasized in \cite{Martelli:2023oqk} and references therein) or the JK‑residue formula (used in \cite{Cassia:2025aus} and references therein). For our purposes, we only need that the generating function is algorithmically computable from a charge matrix $Q^a_i$; we formally adopt the latter approach below to evaluate equivariant characteristic classes as derivatives of $\BV_X$. We further define the $m$-tuple equivariant intersection numbers and the mesonic equivariant volume~\footnote{The name originates from the mesonic twist, i.e., the blow‑down of internal two‑cycles: $t^a = 0$, $\forall a$.} (or zeroth intersection number) as
\be
\label{eq:Cgeometry}
 C^X_{i_1,\dots,i_m} (\e) := \frac{\partial^m \BV_X(\lam,\e)}
 {\partial\lam^{i_1}\cdots\partial\lam^{i_m}} \Big|_{\lam=0}\ , \qquad
	C_X (\e) := \BV_X (\lambda = 0, \epsilon)\ ,
\ee
which can be used to uniquely determine the characteristic numbers of $X$, to which we turn next.~\footnote{A subtle point, of little importance here but relevant for holographic matches (see e.g.\ \cite{Hristov:2026zjh}): equivariant intersection numbers uniquely determine the characteristic numbers, but themselves depend on the choice of JK‑residue chamber (or resolution of $X$).} Note that the equivariant parameters $\e_i$ and their conjugates $\lam^i$ redundantly overparametrize the faithful equivariant parameters (denoted $\nu_a$ and $\mu^\alpha$, respectively, in \cite{Cassia:2025aus}).\\

{\bf Tangent bundle}\\
Consider first the characteristic classes of $X$ itself. By an abuse of notation (which aids the physics picture) we write $c_p(X)$ instead of the more precise $c_p(TX)$. To compute their equivariant generalizations – denoted by a $\mathbb{T}$ superscript following \cite{Martelli:2023oqk} – we insert the equivariant Chern roots $x_i$ into the JK‑residue formula. For example, the $p$-th Chern class corresponds to the replacement
\be
	c_p^\mathbb{T} (X)  \rightarrow \sum_{i_1 < \dots < i_p} x_{i_1} \dots x_{i_p}\ ,
\ee
while the Pontryagin classes follow simply via 
\be
p_1^\mathbb{T} (X) := (c_1^\mathbb{T} (X))^2 - 2 c_2^\mathbb{T} (X)\ , \qquad p_2^\mathbb{T} (X) := (c_2^\mathbb{T} (X))^2-2 c_1^\mathbb{T} (X) c_3 ^\mathbb{T} (X) + 2 c_4^\mathbb{T} (X)\ .
\ee
It follows that the equivariant Chern and Pontryagin numbers of the tangent bundle are given by~\footnote{It might seem counterintuitive at first, but we do not need to integrate only a top equivariant form in order to find a non-vanishing answer.}
\be
\label{eq:shorthandc2andc3}
\begin{split}		
	c^X_p (\e) := \int_X c_p^\mathbb{T} (X) &= \oint_\text{JK} \prod_{a=1}^r\frac{\mathd\phi_a}{2\pi\mathi} \frac{ \sum_{i_1 < \dots < i_p} x_{i_1} \dots x_{i_p}}{\prod_{i=1}^n x_i}=  \sum_{i_1 < \dots < i_p} C^X_{i_1,\dots,i_p}\ , \\
	p^X_1 (\e)  := \int_X p_1^\mathbb{T} (X) &= \dots = \sum_{i, j}  C^X_{i, j} - 2\, \sum_{i<j}  C^X_{i, j} \ , \\
	 p^X_2 (\e) := \int_X p_2^\mathbb{T} (X)&= \dots= \sum_{i<j} \sum_{k<l}  C^X_{i, j, k, l} - 2  \sum_{i} \sum_{j<k<l}C^X_{i, j, k, l} + 2 \sum_{i<j<k<l} C^X_{i, j, k, l} \ ,
\end{split}
\ee
noting that the top equivariant Chern class matches the ordinary non-equivariant one and is a topological invariant independent of $\e$, $\chi (X) := c^X_d (\e)$.

In addition, we use the short-hand notation
\be
\label{eq:defk}
	k^X_p (\e) := \frac1{C_X (\e)}\, c^X_p (\e) = \frac1{C_X (\e)}\, \Big[ \frac{\partial^p \BV_X(\lam,\e)}{\partial\lam^{i_1}\cdots\partial\lam^{i_p}} \Big]_{\lam=0}\ ,
\ee
which allows a more efficient repackaging of the Pontryagin numbers,
\be
	 p^X_1 (\e)= C_X (\e) \left( (k^X_1 (\e))^2 - 2\, k^X_2 (\e) \right)\ , \qquad
p^X_2 (\e) = C_X (\e) \left( (k^X_2 (\e))^2 -2\, k^X_1 (\e) k^X_3 (\e)+2\, k^X_4 (\e) \right)\ .
\ee
These identities follow directly from the linear $\lam$ dependence in the exponent of $\BV_X(\lam,\e)$, c.f.\ \eqref{eq:evol-JK}.\\

{\bf Calabi-Yau condition}\\
 The CY condition for a vanishing (non-equivariant) first Chern class of $X$ can be translated into the following condition on the matrix of charges:
\be
\label{eq:CYcondonQ}
	\sum_i Q^a_i = 0\ , \qquad \forall a\ ,
\ee
which leads to 
\be
\label{eq:a10}
\begin{split}
	c_1^\mathbb{T} (X) \rightarrow \sum_{i=1}^n (\e_i + \phi_a Q^a_i) = \sum_{i=1}^n \e_i\ , \quad
\Rightarrow \quad  c^X_1 (\e) = k^X_1 (\e)\, C_X (\e) = (\sum_{i=1}^n \e_i)\, C_X (\e)\ .
\end{split}
\ee
The expression for $k_1^X$ is universal for any CY manifold. No such simplification exists for $C_X$ and $k_{p \geq 2}^X$; they remain arbitrary rational functions of the $\e$-parameters (with fixed homogeneity degrees $-d$ and $p$, respectively) that depend on $Q^a_i$ via the chain of identities above. Many explicit examples of local toric CY manifolds and calculations of $\BV(\lam,\e)$ can be found in \cite{Cassia:2022lfj,Martelli:2023oqk,Cassia:2025aus,Hristov:2026zjh}. \\

{\bf Normal bundle, $Z = \BC^2$}\\
We also need the characteristic classes of the normal bundle $Z$ (see main text), which corresponds to the $SO(5)$ R‑symmetry bundle of M5‑branes on flat space – equivalently, the symmetry of the four‑sphere in the near‑horizon geometry. Breaking the R‑symmetry to its Cartan subgroup $U(1)^2$, we can take the normal bundle to be $Z = \BC^2$. This was shown in \cite{Martelli:2023oqk}[Section 5.2.4] using the equivariant volume. We denote the two equivariant parameters of $\BC$ (the Chern roots of the $SO(5)$ bundle) by $\Delta_{1,2}$. Their Chern classes give the following insertions in the JK‑residue formula:
\be
\label{eq:a11}
	c_1^\mathbb{T} (Z) \rightarrow \sum_i \Delta_i = \Delta_1 + \Delta_2\ , \qquad c_2^\mathbb{T} (Z) \rightarrow \sum_{i<j} \Delta_i \Delta_j = \Delta_1 \Delta_2\ , \qquad c^\mathbb{T}_{p > 2} (Z) \rightarrow 0\ .
\ee
We have not allowed for non‑trivial magnetic fluxes through compact two‑cycles of $X$; such fluxes would mix the equivariant Chern roots of the normal and tangent bundles, as detailed in \cite{Martelli:2023oqk}[Section 4.2]. The non-vanishing Chern classes of the normal bundle in turn contribute to the Pontryagin classes,
\be
	p^\mathbb{T}_1 (Z) = (c^\mathbb{T}_1 (Z))^2 - 2 c^\mathbb{T}_2 (Z) \rightarrow (\Delta_1)^2 + (\Delta_2)^2\ , \qquad p^\mathbb{T}_2 (Z) = (c^\mathbb{T}_2 (Z))^2 \rightarrow (\Delta_1 \Delta_2)^2\ .
\ee
such that the equivariant integrals appearing in the main text are simply evaluated as
\bea
\label{eq:a13}
\begin{split}
	c_1^Z  = \int_{X} c_1^\mathbb{T} (Z) &= (\Delta_1 + \Delta_2)\, C_X (\e)\ , \qquad c_2^Z = \int_{X} c_2^\mathbb{T} (Z) = (\Delta_1\, \Delta_2)\, C_X (\e) = k_2^Z (\Delta)\, C_X (\e)\ ,\\
	\int_{X} (p_1^\mathbb{T} (Z))^2 &= ( (\Delta_1)^2 + (\Delta_2)^2)^2\, C_X (\e)\ , \qquad \int_{X} p_2^\mathbb{T} (Z) = (\Delta_1 \Delta_2)^2\, C_X (\e)\ , \\
	\int_{X} p_1^\mathbb{T} (Z)\, & p_1^\mathbb{T} (X) = ( (\Delta_1)^2 + (\Delta_2)^2)\, \left( (k^X_1 (\e))^2 -2 k^X_2 (\e) \right)\, C_X (\e)\ .
\end{split}
\eea

\end{document}